\documentclass[12pt,cls,onecolumn]{IEEEtran}
\usepackage{graphicx,amsmath,amssymb,epsfig, amsfonts, cite, latexsym, cuted, multicol, multirow, subfigure, stfloats, array, tabularx}
\usepackage{subeqnarray,tabularx}
\usepackage{color}
\usepackage{setspace}
\usepackage{anysize}

\begin{document}

\title{A New Understanding of Friendships in Space: Complex Networks Meet Twitter}
\author{\large Won-Yong Shin, Bikash C. Singh, Jaehee Cho, and
Andr{\'e} M. Everett
\\
\thanks{This research was supported by the Basic Science Research
Program through the National Research Foundation of Korea (NRF)
funded by the Ministry of Education (2014R1A1A2054577).}
\thanks{W.-Y. Shin (corresponding author) and B. C. Singh are with the Department of Computer Science and
Engineering, Dankook University, Yongin 448-701, Republic of Korea
(E-mail: wyshin@dankook.ac.kr; bikash070@gmail.com).}
\thanks{J. Cho is with the Department of Business Administration, Kwangwoon University, Seoul 139-701, Republic of Korea (E-mail: mis1@kw.ac.kr)}
\thanks{A. M. Everett is with the Department of Management, University of Otago,
Dunedin 9054, New Zealand (E-mail: andre.everett@otago.ac.nz).}
} \maketitle


\markboth{Journal of Information Science} {Shin {\em et al.}: A
New Understanding of Friendships in Space: Complex Networks Meet
Twitter}


\newtheorem{definition}{Definition}
\newtheorem{theorem}{Theorem}
\newtheorem{lemma}{Lemma}
\newtheorem{example}{Example}
\newtheorem{corollary}{Corollary}
\newtheorem{proposition}{Proposition}
\newtheorem{conjecture}{Conjecture}
\newtheorem{remark}{Remark}

\def \diag{\operatornamewithlimits{diag}}
\def \min{\operatornamewithlimits{min}}
\def \max{\operatornamewithlimits{max}}
\def \log{\operatorname{log}}
\def \max{\operatorname{max}}
\def \rank{\operatorname{rank}}
\def \out{\operatorname{out}}
\def \exp{\operatorname{exp}}
\def \arg{\operatorname{arg}}
\def \E{\operatorname{E}}
\def \tr{\operatorname{tr}}
\def \SNR{\operatorname{SNR}}
\def \dB{\operatorname{dB}}
\def \ln{\operatorname{ln}}

\def \bmat{ \begin{bmatrix} }
\def \emat{ \end{bmatrix} }

\def \be {\begin{eqnarray}}
\def \ee {\end{eqnarray}}
\def \ben {\begin{eqnarray*}}
\def \een {\end{eqnarray*}}

\begin{abstract}
Studies on friendships in online social networks involving
geographic distance have so far relied on the city location
provided in users' profiles. Consequently, most of the research on
friendships have provided accuracy at the {\em city level}, at
best, to designate a user's location. This study analyzes a
Twitter dataset because it provides the exact geographic distance
between corresponding users. We start by introducing a strong
definition of ``{\em friend}" on Twitter (i.e., a definition of
{\em bidirectional friendship}), requiring bidirectional
communication. Next, we utilize {\em geo-tagged mentions}
delivered by users to determine their locations, where
``@username" is contained anywhere in the body of tweets. To
provide analysis results, we first introduce a friend counting
algorithm. From the fact that Twitter users are likely to post
consecutive tweets in the static mode, we also introduce a
two-stage distance estimation algorithm. As the first of our main
contributions, we verify that the number of friends of a
particular Twitter user follows a well-known power-law
distribution (i.e., a Zipf's distribution or a Pareto
distribution). Our study also provides the following
newly-discovered friendship degree related to the issue of space:
The number of friends according to distance follows a {\em double
power-law} (i.e., a {\em double Pareto law}) distribution,
indicating that the probability of befriending a particular
Twitter user is significantly reduced beyond a certain geographic
distance between users, termed the {\em separation point}. Our
analysis provides concrete evidence that Twitter can be a useful
platform for assigning a more accurate scalar value to the degree
of friendship between two users.
\end{abstract}


\newpage

\section{Introduction}

In recent years, research in the field of online social networks
(OSNs) has grown dramatically with the evolution of technologies
while harnessing Big Data. Focusing on the relationships (edges)
among users or profiles (vertices), OSN analysis has emerged as
one of the most popular and familiar approaches for examining
interaction, information sharing, and collaboration among online
users~\cite{Wilson:09}. Simultaneously, the field of {\em complex
networks} has emerged as an independent research area, with strong
connections to random graph theory from mathematics as well as to
social network analysis by physicists, interested in understanding
the behaviors of large-scale interacting networks. Based on
massive datasets of large-scale real-world OSNs such as
Twitter~\cite{Kwak:10}, Facebook~\cite{Viswanath:09},
Flickr~\cite{Mislove:08}, and Foursquare~\cite{Chen:14}, extensive
studies have validated that the small-world phenomenon (originally
introduced by Watts and Strogatz~\cite{Watts:98}) and scale-free
degree distribution,\footnote{A ``small-world" network is a type
of mathematical graph in which two arbitrary nodes (people) are
connected by a short chain of intermediate links (friends), and a
``scale-free" network is a network whose degree distribution
follows a power-law.} which are the two most representative
features of complex networks, nearly hold in
OSNs~\cite{Svenson:06}. Twitter is one of the most popular
micro-blogs (or social media), allowing users to ``tweet" about
any topic within the 140-character limit and to ``follow" others
to receive their tweets. At the start of 2015, Twitter played a
vital role in facilitating social contacts, boasting 284 million
active users per month, publishing 500 million tweets daily from
their web browsers and smart
phones.\footnote{https://about.twitter.com/company}

\subsection{Related Work}

To understand the nature of friendships online with respect to
geographic distance, some efforts have focused on users' online
profiles that include their city of
residence~\cite{Liben-Nowell:05,Kaltenbrunner:12}.
In~\cite{Liben-Nowell:05}, experimental results based on the
LiveJournal social network\footnote{https://www.livejournal.com}
demonstrated a close relationship between geographic distance and
probability distribution of friendship, where the probability of
befriending a particular user on LiveJournal is inversely
proportional to the positive power of the number of closer users.
Contrary to~\cite{Liben-Nowell:05}, based on the data collected
from Tuenti,\footnote{https://www.tuenti.com} a Spanish social
networking service, it was found in~\cite{Kaltenbrunner:12} that
social interactions online are only weakly affected by spatial
proximity, with other factors dominating.

However, the effect of distance on online social interactions has
not yet been fully understood. In the previous studies, the
geographic location points only to the location of users at a {\em
city scale}. For this reason, the friendship degree distribution
contains a background probability that is independent of geography
due to the city-scale
resolution~\cite{Liben-Nowell:05,Kaltenbrunner:12}. On the other
hand, {\em geo-located Twitter} can provide high-precision
location information down to 10 meters through the Global
Positioning System (GPS) interface~\cite{Jurdak:13} of users'
smart phones while offering comprehensive metadata with a gigantic
sample of the whole population.

For this reason, there is extensive and growing interest among
researchers to understand a variety of social behaviors through
geo-located Twitter, or, equivalently, geo-tagged
tweets~\cite{Takhteyev:12,Kulshrestha:12,Frias-Martinez:12,Alowibdi:14,Lee:10,Hawelka:14,Jurdak,Liu,Falcone:14}.
Even if geo-tagged tweets account for approximately 1\% of the
total amount~\cite{Morstatter:13}, thanks to the increasing
penetration of smart devices and mobile applications, the volume
of geo-located Twitter has grown constantly and now forms an
invaluable register for understanding human behavior and modelling
the way people interact in space. In~\cite{Takhteyev:12}, along
with geo-locations for collected tweets, analysis included how
geo-related factors such as physical distance, frequency of air
travel, national boundaries, and language differences affect
formation of social ties on Twitter. In~\cite{Kulshrestha:12}, it
was found that the geo-locations of Twitter users across different
countries considerably impact their participation in Twitter,
their connectivity with other users, and the information that they
exchange with each other. As another application, the use of
geo-tagged tweets was evaluated as a complementary source of
information for urban planning including i) a technique to
determine land uses in a specific urban area based on tweeting
patterns and ii) a technique to identify urban points of interest
at places with high tweeting activity~\cite{Frias-Martinez:12}.
New approaches based on geo-tagged tweets were also proposed to
find top vacation spots for a particular holiday by applying
indexing, spatio-temporal querying, and machine learning
techniques~\cite{Alowibdi:14} and to detect unusual geo-social
events by measuring geographical regularities of crowd
behaviors~\cite{Lee:10}.

Benefiting from the increasing availability of location
information from geo-tagged tweets, there has been a steady push
to understand individual human
mobility~\cite{Hawelka:14,Jurdak,Liu,Falcone:14}, which is of
fundamental importance for many applications to human and
electronic virus prediction and traffic and population
forecasting. Recent effort has focused on the studies of human
mobility using tracking technologies such as mobile
phones~\cite{Gonzalez:08,Song:10,Jiang:13,Wang:11}, GPS
receivers~\cite{Rhee:11}, WiFi logging~\cite{Chaintreau:07},
Bluetooth~\cite{Hui:08}, and RFID devices~\cite{Cattuto:10} as
well as location-based social network check-in data~\cite{Cho:11},
but these technologies involve privacy concerns or data access
restrictions. In contrast, geo-tagged tweets can capture much
richer features of human mobility. For example,
in~\cite{Hawelka:14}, global human mobility patterns were widely
revealed, and a comparative study on the mobility characteristics
of different countries was conducted. Furthermore, it was found
in~\cite{Jurdak} that the geo-located Twitter data for Australia
reveals multiple modes of human mobility from intra-site to
metropolitan and inter-city movements. As another point of view,
in~\cite{Liu}, it was reported that in Australia, the gravity law
is applicable for estimating human mobility by showing that
mobility between an origin and its destination is proportional to
the product of populations of these two places and is inversely
proportional to the power-law of distance between them.
In~\cite{Falcone:14}, the problem of labelling the places of a
city based on collected spatio-temporal data was addressed,
including i) to infer whether a place belongs to a certain
category or not and ii) to choose the category of a place among a
set of categories.

\subsection{Main Contributions}

In our work, we utilize {\em geo-tagged mentions} on Twitter, sent
by users, to identify their exact location information. A
`mention' in Twitter consists of inclusion of ``@username"
anywhere in the body of tweets. From the fact that we tend to
interact offline with people living very near to us, we derive as
a natural extension the question whether geography and social
relationships are inextricably intertwined on Twitter. Our
research significantly differs from a variety of studies on human
mobility in the
literature~\cite{Hawelka:14,Jurdak,Liu,Falcone:14,Gonzalez:08,Song:10,Jiang:13,Wang:11,Rhee:11,Chaintreau:07,Hui:08,Cattuto:10,Cho:11}
since it is interested in how a pair of users interacts. To the
best of our knowledge, such an attempt to analyze one-to-one
friendship based on geo-located tweets (or mentions) has not yet
been described in the literature.

As people normally spend a substantial amount of time online, data
regarding these two dimensions (i.e., geography and online social
relationships) are becoming increasingly precise, thus motivating
us to build more reliable models to describe social
interactions~\cite{Backstrom:10}. Previous studies have employed
large amounts of data from diverse sources, such as smart devices
and web-based applications, to examine how social data resources
(e.g., photos on Flickr) are processed with
tagging~\cite{Giannakidou:08,Nguyen:15}. Both a co-clustering
approach~\cite{Giannakidou:08} and a spatial ranking
approach~\cite{Nguyen:15} have been introduced to discover
meaningful relationships between a set of relevant resources and a
set of tags. This paper goes beyond past research to determine how
friendship patterns are geographically represented by Twitter,
analyzing a single-source dataset (to avoid potential confounds)
that contains a huge number of geo-tagged mentions from users in
i) the state of California in the United States (US) and Los
Angeles (the most populous city in the state) and ii) the United
Kingdom (UK) and London (the most populous city in the UK). These
two location sets were selected as demographically comparable, yet
distinct and geographically separated, leading adopters of Twitter
with sufficient data to enable meaningful comparative analysis for
our intentionally exploratory study (which will be specified in
Section II). In this dataset, each mention record has a geo-tag
(spatial information) and a timestamp (temporal information)
indicating from where, when, and by whom the mention was sent. We
propose and apply the following new framework, which establishes a
more accurate friendship degree on Twitter, and a method to enable
analysis based on geographic distance:

\begin{itemize}
\item To fully take into account the intensity of communication
between users, we start our analysis by introducing a rather
strong definition of ``{\em friend}" on Twitter, i.e., a
definition of {\em bidirectional friendship}, instead of
na\"{\i}vely considering the set of followers and followees
(unidirectional terms). This definition requires bidirectional
communication within a designated time frame to constitute a
friendship.

\item Using the above definition, we introduce a friend counting
algorithm, which computes the distribution of the number of
friends for each Twitter user.

\item By showing that almost all Twitter users are likely to post
consecutive tweets in the static mode, we propose a two-stage
distance estimation method, where the geographic distance between
two befriended users (denoted by Users $u$ and $v$) based on our
definition of bidirectional friendship is estimated by
sequentially measuring the two senders' locations. More
specifically, the location of User $u$ is recorded at the moment
when User $u$ sends a mention to User $v$, while the location of
User $v$ can also be recorded when User $v$ sends a replied
mention to User $u$ at the next closest time, enabling estimation
of the distance between Users $u$ and $v$.
\end{itemize}

Note that the above definition is suitable for evaluating
one-to-one bidirectional social interactions on Twitter since
Twitter users tend to personally interact with only a few of their
followers/followees by sending and receiving direct mentions. We
would like to synthetically analyze how the geographic distance
between Twitter users affects their interaction, based on our new
framework. Our main contributions are as follows:

\begin{itemize}
\item Based on the definition of bidirectional friendship, we
first verify that the number of friends of {\em one user} on
Twitter follows a power-law distribution (i.e., a Zipf's
distribution~\cite{Manning:99} or a Pareto
distribution~\cite{Newman:05}) even on Twitter, which is known to
be asymptotically equivalent to the degree distribution of
scale-free networks. This finding is consistent with the earlier
results in other OSNs.

\item Next, more interestingly, we characterize a newly-discovered
probability distribution of the number of friends according to
{\em geographic distance}, which does not follow a homogeneous
power-law but, instead, a {\em double power-law} (i.e., a {\em
double Pareto law}~\cite{Reed:03}). From this new finding, we
identify not only two fundamentally {\em separate regimes}, termed
the intra-city and inter-city regimes, which are characterized by
two different power-laws in the distribution, but also the {\em
separation point} between these regimes.
\end{itemize}

\subsection{Organization}

The rest of this paper is organized as follows. Section II
describes the dataset, and Section III explains our analysis
methodology. In Section IV, experimental results are presented by
analyzing the number of friends of a particular user and the
number of friends with respect to distance. Finally, we summarize
the paper with some concluding remarks in Section V.

\section{Dataset} \label{SEC:dataset}

We use a dataset collected from crawling the Twitter network via
Twitter Streaming Application Programming Interface
(API),\footnote{https://dev.twitter.com/decs/streaming-apis} which
returns tweets matching a query provided by the Streaming API
user. Although the Twitter Streaming API only returns at most a
1\% sample of all the tweets produced at a given moment, it
constitutes a valid representation of users' activity on Twitter
when more specific parameter sets such as different users,
geographic bounding boxes, and keywords are created (thereby
enabling extraction of more data from the Streaming
API)~\cite{Morstatter:13,Morstatter:14}. It was found that the
Streaming API returns an almost complete set of {\em geo-tagged}
tweets despite sampling~\cite{Morstatter:13}. Thus, there is no
doubt that this research is working with an almost complete sample
of geo-located Twitter data.

In our work, we examined data from all possible devices (sources)
that indicate the user's location information at the time that
they access Twitter. The statistics based on our dataset
demonstrate that a large majority of the Twitter users in our
sample posted geo-tagged tweets through smart phones rather than
web browsers on a desktop or laptop computer.\footnote{We note
that smart devices and mobile applications enable us to provide
high-precision location information through the built-in GPS
interface. On the other hand, with the Geo-location API, web
browsers can detect the users' approximate location information
inferred from network signals such as IP address, WiFi, Bluetooth,
MAC address, and GSM/CDMA cell ID, which are not guaranteed to
return the users' actual location. Based on our dataset, it is
found that 77.84\% and 82.21\% of Twitter users tend to post
geo-tagged tweets in California and the UK, respectively, via
iPhone and Android Phone, which are the smart phone types using
the two most popular mobile platforms among all devices. It is
also found that 90.52\% and 81.14\% of posted geo-tagged tweets
tend to be recorded in California and the UK, respectively, via
iPhone and Android Phone.} This reveals that our dataset is much
more inclined toward geo-tagged tweets (more rigorously,
geo-tagged mentions) transmitted through the GPS interface.

The dataset consists of a huge amount of geo-tagged mentions
recorded from Twitter users from September 22, 2014 to October 23,
2014 (about one month) in the following four large regions:
California, Los Angeles, the UK, and London. Note that this
short-term (one month) dataset is sufficient to examine how
closely one user has recently interacted with another online
(i.e., a personal online relationship between two users). The four
regions in our dataset were selected since they are quite
comparable at both the macro (state or country) and micro (city)
scales in terms of i) area, ii) population density, and iii)
Twitter popularity (e.g., the number of Twitter accounts or the
number of posted tweets). The comparison between location sets for
the aforementioned three representative attributes is summarized
in TABLE~I, divided according to the types of two geographic
scales.\footnote{http://en.wikipedia.org/wiki/California\\
http://en.wikipedia.org/wiki/United\_Kingdom\\
http://en.wikipedia.org/wiki/Los\_Angeles\\
http://en.wikipedia.org/wiki/London\\
http://semiocast.com/publications/2012\_07\_30\_Twitter\_reaches\_half\_a\_billion\_accounts\_140m\_in\_the\_US}

\begin{table}[t!] \label{TABLE1}
\renewcommand{\arraystretch}{1.1}
\centering{%
\caption{Comparison of the location sets.} \subfigure[California
versus UK (state scale or
country scale)]{%
 \begin{tabular}{|c||c|c|}
   \hline
    Attribute & California & UK   \\
 \hline
 \hline
    Area (km$^2$) & 423,970 & 243,610   \\
 \hline
   Population density (population/km$^2$) & 95.0 & 225.6  \\
 \hline
   Global ranking among countries & 1st (US as whole country) & 4th \\

    by the number of Twitter accounts & &   \\
   \hline
\end{tabular}
\label{TABLE1a} } \vspace{0.3cm}

\subfigure[Los Angeles versus London (city scale)]{%
 \begin{tabular}{|c||c|c|}
   \hline
    Attribute & Los Angeles & London   \\
 \hline
 \hline
    Area (km$^2$) & 1,302 & 1,572   \\
 \hline
   Population density (population/km$^2$) & 3,198 & 5,354  \\
 \hline
   Global ranking among cities & 8th & 3rd \\

    by the number of posted tweets (June 2012) & &   \\
   \hline
\end{tabular} \label{TABLE1b}}
} \vspace{0.6cm}
\end{table}


%
%
%
%
%
%

\begin{table}[t!]
\renewcommand{\arraystretch}{1.1}
\begin{center}
\caption{Statistics of the dataset: The number of mentions and
unique users in each region.}

 \begin{tabular}{|c||c|c|}
   \hline
    Region & Number of mentions & Number of users (senders)   \\
 \hline
 \hline
    California & 2,349,901 & 217,439   \\
 \hline
  Los Angeles & 918,360 & 51,625  \\
 \hline
   UK & 3,721,716 & 612,368  \\
   \hline
   London & 614,045 & 58,046 \\
   \hline
\end{tabular}
\label{TABLE2}
\end{center}

\end{table}

The representative statistics of the collected dataset, such as
the total number of mentions and the total number of senders, are
also summarized by regional group in TABLE~II. In this dataset,
each mention record has a geo-tag and a timestamp indicating from
where, when, and by whom the mention was sent. Based on this
information, we are able to construct a user's location history
denoted by a sequence $L=(x_{ki},y_{ki},t_i)$, where $x_{ki}$ and
$y_{ki}$ are the $x$- and $y$-coordinates of User $k$ at time
$t_i$, respectively. The location information provided by the
geo-tag is denoted by latitude and longitude, which are measured
in degrees, minutes, and seconds.

Each mention on Twitter contains a number of entities that are
distinguished by their attributed fields. For data analysis, we
adopted the following five essential fields from the metadata of
mentions:\footnote{https://dev.twitter.com/overview/api/tweets}

\begin{itemize}
\item {\em user\_id\_str}: string representation of the sender ID

\item {\em in\_reply\_to\_user\_id\_str}: string representation of
the receiver ID

\item {\em lat}: latitude of the sender

\item {\em lon}: longitude of the sender

\item {\em created\_at}: UTC/GMT time when the mention is
delivered, i.e., the timestamp
\end{itemize}

Note that the two location fields, {\em lat} and {\em lon},
corresponds to spatial (geo-tagged) information while the last
field, {\em created\_at}, represents temporal (time-stamped)
information.

\section{Research Methodology}

We start by introducing the following definition of
``bidirectional friendship" on Twitter.

\begin{definition}[Bidirectional friendship in Twitter]
If two users send/receive direct mentions to/from each other
(i.e., bidirectional personal communication occurs) within a
designated amount of time, then they form a bidirectional
friendship with each other.
\end{definition}

Note that our definition differs from the conventional definition
of ``friend" on Twitter, which is referred to as a followee and
thus represents a {\em unidirectional}
relation~\cite{Hodas:13,Bastos:12}.\footnote{Twitter shows a low
level of reciprocity; 77.9\% of user pairs with any link between a
Twitter user and his/her follower are connected one-way, and only
22.1\% exhibit a reciprocal relationship between them (i.e.,
two-way links)~\cite{Kwak:10}.} Since friendship relations in the
offline world and on other OSNs such as Facebook~\cite{Ugander}
are generally not unidirectional, our intention is to formulate a
{\em bidirectional} friendship that can be directly applicable to
offline relationships. This strong definition enables exclusion of
{\em inactive friends} (or passive friends) who have been out of
contact online for a long designated amount of time (e.g., about
one month in our work) and to count the number of {\em active
friends} who have recently communicated with each other.

\subsection{Counting Number of Friends of a Particular User}

\begin{figure}[t]
  \begin{center}
  \leavevmode \epsfxsize=0.45\textwidth
  \epsffile{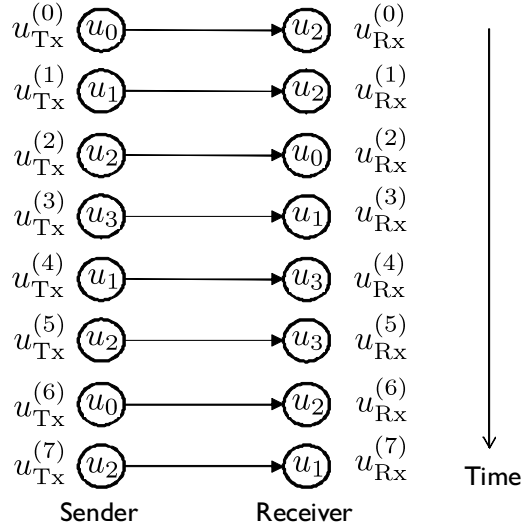}
  \caption{One example that illustrates how geo-tagged mentions are delivered from senders to receivers according to time sequence, where $u_{\text{Tx}}^{(t)}$ and $u_{\text{Rx}}^{(t)}$ denote the transmitter and the corresponding receiver at time $t\in\{0,1,\cdots\}$. In this example, three pairs of friends, $(u_0,u_2)$, $(u_1,u_2)$, and $(u_1,u_3)$, are made among four users $u_0$, $u_1$, $u_2$, and $u_3$.}
  \label{FIG:friends}
  \end{center}
\end{figure}

In this subsection, we explain how to count the number of friends
of each user who sent at least one geo-tagged mention. Suppose
that there are four Twitter users, denoted by $u_0$, $u_1$, $u_2$,
and $u_3$, who sent or received at least one geo-tagged mention
according to temporal event sequences, as illustrated in
Figure~\ref{FIG:friends}. Here, $u_{\text{Tx}}^{(t)}$ and
$u_{\text{Rx}}^{(t)}$ denote the transmitter and the corresponding
receiver sequentially at time instance $t\in\{0,1,\cdots\}$. In
this example, according to the aforementioned definition, three
pairs of friends $(u_0,u_2)$, $(u_1,u_2)$, and $(u_1,u_3)$ are
found out of the above user set. Moreover, one can find that the
number of friends of each user $u_0$, $u_1$, $u_2$, and $u_3$ is
given by 1, 2, 2, and 1, respectively. In our framework, if
bidirectional communication between two certain users occurs at
least once, then their friendship degree is set to one. Otherwise,
it is set to zero, i.e., no friendship between the two users is
created. That is, even with more than two bidirectional
communications between two users, their friendship degree is
maintained at one in this binary or Boolean evaluation. In our
sample space, we exclude the user set whose friendship degree is
zero since including such users will lead to scaling down the
probability distribution of the nonzero number of friends.

The overall procedure of the friend counting algorithm ({\bf
Algorithm 1}) is described in TABLE~\ref{algorithm1}, where
$n_{u}$ denotes the number of friends of User
$u\in\{u_0,u_1,\cdots,u_{I-1}\}$ who sent a geo-tagged mention to
User $v\in\{v_0,v_1,\cdots,v_{J-1}\}$, and $I$ and $J$ are the
total number of senders and receivers in a dataset, respectively.

\begin{table}[!t]
\renewcommand{\arraystretch}{1.1}
 \centering \caption{The overall procedure of the
friend counting algorithm.}
\begin{tabular}{l}
\hline
\textbf{Algorithm 1} Friend counting algorithm\\
\hline \textbf{Input}: $u_{\text{Tx}}^{(t)}$ and
$u_{\text{Rx}}^{(t)}$ for $t=0,1,\cdots,T-1$,
$u\in\{u_0,u_1,\cdots,u_{I-1}\}$ \\and
$v\in\{v_0,v_1,\cdots,v_{J-1}\}$\\
\textbf{Output}: $n_{u}$ for all $u$\\
\textbf{Initialization}: $c_{uv}\leftarrow0$ and
$n_{u}\leftarrow0$
for all $u$ and $v$\\
00: \textbf{for} $t\leftarrow0$ \textbf{to} $T-1$ \textbf{do}
\\
01: \hspace{0.3cm} Find the user indices $u$ and $v$ for
$u_{\text{Tx}}^{(t)}$ and $u_{\text{Rx}}^{(t)}$, respectively\\
02: \hspace{0.3cm} \textbf{for} $s\leftarrow t+1$ \textbf{to}
$T-1$ \textbf{do}\\
03: \hspace{0.6cm} \textbf{if}
($u_{\text{Tx}}^{(s)}==u_{\text{Rx}}^{(t)}$) \textbf{then}\\
04: \hspace{0.9cm} \textbf{if}
($u_{\text{Rx}}^{(s)}==u_{\text{Tx}}^{(t)}$) \textbf{then}\\
05: \hspace{1.2cm} $c_{uv}\leftarrow 1$ \\
06: \hspace{1.2cm} break (go back to line 00) \\
07: \hspace{0.9cm} \textbf{end if} \\
08: \hspace{0.6cm} \textbf{end if} \\
09: \hspace{0.3cm} \textbf{end for}\\
10: \textbf{end for}\\
11: \textbf{for} all $u$ and $v$ \textbf{do} \\
12: \hspace{0.3cm} $n_{u}\leftarrow n_{u} + c_{uv}$ \\
13: \textbf{end for} \\
 \hline
\end{tabular} \label{algorithm1}
\end{table}

\subsection{Finding Friend Distribution With Respect to Distance}

In this subsection, let us turn to characterizing the friendship
degree of individuals regarding geography by analyzing their
sequences $L=(x_{ui},y_{ui},t_i)$ of geo-tagged mentions, where
only the senders' location information is recorded. We propose a
two-stage method to estimate the geographic distance between
Twitter friends. If User $u$ sends a mention to User $v$, then the
location information of User $u$ is recorded (the first stage). In
order to find the location of User $v$, we need to wait for the
moment at which User $v$ sends a mention back to User $u$ (the
second stage). That is, after bidirectional communication between
two Twitter users occurs, the location of each user can be
identified.

It is not possible to evaluate the geographic distance between two
Twitter users through a one-shot process due to the fact that the
location information of only the sender is recorded at a given
instance when a geo-tagged mention is sent. Moreover, because of
the users' movements, it is, however, not straightforward to
measure the exact distance. In this subsection, we introduce a
two-stage distance estimation method, where the geographic
distance between two befriended users is estimated by sequentially
measuring the two senders' locations.

\begin{figure}[t]
  \begin{center}
  \leavevmode \epsfxsize=0.54\textwidth
  \epsffile{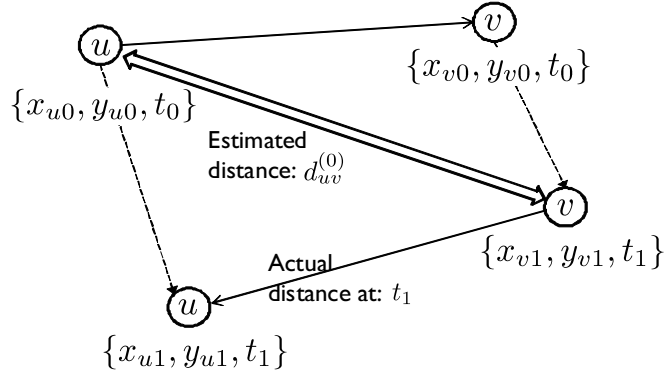}
  \caption{User movement in which User $k\in\{u,v\}$ changes location from $(x_{k0},y_{k0},t_0)$ to $(x_{k1},y_{k1},t_1)$ between sending a geo-tagged mention and receiving a corresponding replied mention.}
  \label{FIG:movement}
  \end{center}
\end{figure}

Before describing the estimation algorithm, let us first focus on
the time interval between the following two events for a
befriended pair: a mention and its {\em replied} mention at the
next closest time. We count only the events with a time duration
between a mention and its replied mention, or inter-mention
interval, of {\em less than one hour} to exclude certain
inaccurate location information that may occur due to users'
movements.\footnote{Note that inter-mention interval of one hour
may be shortened, but this will lead to a reduction in the
available dataset.} Figure~\ref{FIG:movement} illustrates the
instance for which User $u$, originally placed at
$(x_{u0},y_{u0},t_0)$, sent a mention to User $v$ at
$(x_{v0},y_{v0},t_0)$, and then received a replied mention at the
location $(x_{u1},y_{u1},t_1)$ from User $v$ placed at
$(x_{v1},y_{v1},t_1)$. Here, the single solid arrows indicate the
actual distances at time instances $t_0$ and $t_1$ while the
double solid arrow indicates the estimated distance. The distance
that users moved between the two moments in time $t_0$ and $t_1$
(i.e., inter-mention interval) is indicated as the dashed arrows
in the figure. From these two consecutive mention events, it is
possible to estimate the geographic distance based on the two
sequences $(x_{u0},y_{u0},t_0)$ and $(x_{v1},y_{v1},t_1)$. In our
framework, by assuming that the Earth is spherical, we deal with
the shortest path between two users' locations measured along the
surface of the Earth, instead of the rather na\"{\i}ve
straight-line Euclidean distance. Following an approach similar to
that employed in~\cite{Huang:13,Ennis:13}, the distance between
two locations on the Earth's surface can be computed according to
the spherical law of cosines.\footnote{When Sinnott published the
haversine formula~\cite{Sinnott:84}, computational precision was
limited. Nowadays, JavaScript (and most modern computers and
languages) uses IEEE 754 64-bit floating-point numbers, which
provide 15 significant digits of precision. With this precision,
the simple spherical law of cosines formula gives well-conditioned
results down to distances as small as around 1 meter. In view of
this, it is probably worth, in most situations, using the simpler
law of cosines in preference to the haversine formula.} Then, when
we denote the distance between the two users measured from
$(x_{u0},y_{u0},t_0)$ and $(x_{v1},y_{v1},t_1)$ by $d_{uv}^{(0)}$,
we
obtain\footnote{http://mathworld.wolfram.com/SphericalTrigonometry.html}

\begin{align}
d_{uv}^{(0)}=R\cos^{-1}\left(\sin x_{u0} \sin x_{v1} +\cos
x_{u0}\cos x_{v1} \cos \left(y_{v1}-y_{u0}\right) \right),
\label{EQ:distance}
\end{align}
where $R$ [in kilometers (km)] denotes the Earth's radius and is
given as 6,371, and the superscript 0 in $d_{uv}^{(0)}$ represents
the time slot. Here, for notational convenience, it is assumed
that the $x$- and $y$-coordinates represent the latitude and
longitude, respectively.

While the estimated distance (double solid arrow in
Figure~\ref{FIG:movement}) may differ from the actual distance
(single solid arrow in Figure~\ref{FIG:movement}) between Users
$u$ and $v$ at time $t_1$, it is worth noting that people tend to
send/receive multiple consecutive tweets from the same location to
convey a series of ideas~\cite{Jurdak,Liu}. To validate this user
mobility argument, we turn our attention to analyze the
distribution of the number of tweets (i.e., the tweet frequency)
with respect to user velocity.

In our experiments, we use the same dataset collected from the
Twitter users as shown in Section~\ref{SEC:dataset}, but focus on
the two populous metropolitan areas, Los Angeles and London. To
exclude certain inaccurate location information that may exist due
to users' movements, we take into account only the case only where
two consecutive geo-tagged tweet events occur {\em within one
hour}. When the location history for two consecutive geo-tagged
tweets of User $k$ at time slots $t_i$ and $t_{i+1}$ is expressed
as sequences $(x_{ki},y_{ki},t_i)$ and
$(x_{k(i+1)},y_{k(i+1)},t_{i+1})$, respectively, the average
velocity $v_k^{(i)}$ of the user within this time interval is
given by $v_k^{(i)}=d_k^{(i)}/(t_{i+1}-t_i)$, where $d_k^{(i)}$ is
the distance that User $k$ moved during the interval
$[t_i,t_{i+1}]$ and thus is given by $d_k^{(i)}=R\cos^{-1}
\left(\sin x_{ki}\sin x_{k(i+1)} + \cos x_{ki} \cos x_{k(i+1)}
\cos \left(y_{k(i+1)}-y_{ki}\right)\right)$ (refer to equation
(\ref{EQ:distance}) for more details). From the set of average
velocities $\left\{v_k^{(0)},v_k^{(1)},\cdots,v_k^{(T-1)}\right\}$
obtained from all users in the dataset, the tweet frequency can be
categorized according to the user velocity.

\begin{figure}[t]
  \begin{center}
  \leavevmode \epsfxsize=0.49\textwidth
  \epsffile{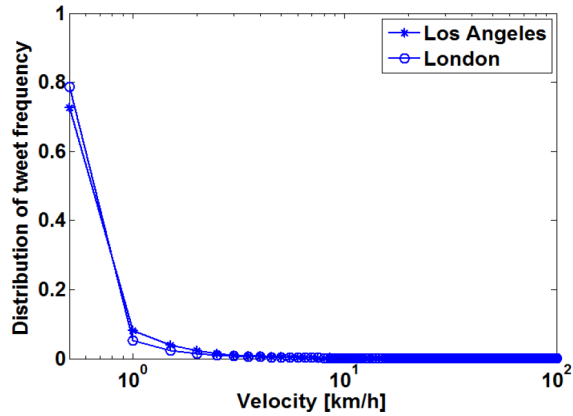}
  \caption{Probability distribution of the tweet frequency with respect to user velocity (log-linear plot).}
  \label{FIG:velocity}
  \end{center}
\end{figure}

Figure~\ref{FIG:velocity} shows the log-linear plot of the
distribution of the number of tweets (i.e., the tweet frequency)
versus the user velocity [km/h], which is obtained from empirical
data. As illustrated in Figure~\ref{FIG:velocity}, most of the
Twitter users (approximately 90\%) in the two metropolitan areas
are likely to post consecutive tweets in the {\em static} mode
whose average velocity ranges from 0 to 2 km/h. Our experiments
also demonstrate that Twitter users in large scale geographic
areas (e.g., state scale (California) or country scale (the UK))
are more likely to post consecutive tweets in the static mode than
city-scale users, even if the results are not presented in
Figure~\ref{FIG:velocity}. Although the inter-tweet interval may
show a different pattern from that of the inter-mention interval
(i.e., the time duration between a mention and its replied mention
from another user), we believe that the above results are
sufficient to support our analysis methodology.

\begin{table}[!t]
\renewcommand{\arraystretch}{1.1}
 \centering \caption{The overall procedure of
distance estimation algorithm.}
\begin{tabular}{l}
\hline
\textbf{Algorithm 2} Distance estimation algorithm\\
\hline \textbf{Input}: $u_{\text{Tx}}^{(t)}$ and
$u_{\text{Rx}}^{(t)}$ for $t=0,1,\cdots,T-1$,
$u\in\{u_0,u_1,\cdots,u_{I-1}\}$ \\and
$v\in\{v_0,v_1,\cdots,v_{J-1}\}$\\
\textbf{Output}: $d_{uv}$ for all $u$ and $v$\\
\textbf{Initialization}: $c_{uv}^{(t)}\leftarrow0$ and
$d_{uv}\leftarrow0$
for all $u$ and $v$\\
00: \textbf{for} $t\leftarrow0$ \textbf{to} $T-1$ \textbf{do}
\\
01: \hspace{0.3cm} Find the user indices $u$ and $v$ for
$u_{\text{Tx}}^{(t)}$ and $u_{\text{Rx}}^{(t)}$, respectively\\
02: \hspace{0.3cm} \textbf{for} $s\leftarrow t+1$ \textbf{to}
$T-1$ \textbf{do}\\
03: \hspace{0.6cm} \textbf{if}
($u_{\text{Tx}}^{(s)}==u_{\text{Rx}}^{(t)}$) \textbf{then}\\
04: \hspace{0.9cm} \textbf{if}
($u_{\text{Rx}}^{(s)}==u_{\text{Tx}}^{(t)}$) \textbf{then}\\
05: \hspace{1.2cm} \textbf{if} (time interval between $t$ and
$s<1$ hour) \textbf{then}\\
06: \hspace{1.5cm} Compute $d_{uv}^{(c_{uv}^{\left(t\right)})}$ in
equation
(\ref{EQ:distance}) \\
07: \hspace{1.5cm} $c_{uv}^{(t)}\leftarrow c_{uv}^{(t)}+1$ \\
08: \hspace{1.5cm} break (go back to line 00) \\
09: \hspace{1.2cm} \textbf{end if} \\
10: \hspace{0.9cm} \textbf{end if} \\
11: \hspace{0.6cm} \textbf{end if} \\
12: \hspace{0.3cm} \textbf{end for} \\
13: \textbf{end for} \\
14: \textbf{for} all $u$ and $v$ \textbf{do} \\
15: \hspace{0.3cm} \textbf{for} $l\leftarrow0$ \textbf{to}
$c_{uv}^{(t)}$ \textbf{do} \\
16: \hspace{0.6cm} $d_{uv} \leftarrow d_{uv} +
d_{uv}^{(l)}/c_{uv}^{(t)}$ \\
17: \hspace{0.3cm} \textbf{end for} \\
18: \textbf{end for} \\
 \hline
\end{tabular} \label{algorithm2}
\end{table}

Now, we are ready to present our distance estimation algorithm
({\bf Algorithm 2}). The overall procedure of the proposed
algorithm is described in TABLE~\ref{algorithm2}, where $d_{uv}$
denotes the estimated geographic distance between user pair
$u\in\{u_0,u_1,\cdots,u_{I-1}\}$ and
$v\in\{v_0,v_1,\cdots,v_{J-1}\}$, and $I$ and $J$ are the total
number of senders and receivers in a dataset, respectively. Note
that as shown in lines 14--18 of the table, the estimated distance
for one pair is obtained by taking the average of all distance
values computed over the available inter-mention intervals, each
of which is less than one hour.

\section{Analysis Results}

In this section, we first verify whether a Zipf's power-law holds
for the Twitter network along with the definition of bidirectional
friendship. Next, we show a newly-discovered distribution of the
number of friends with respect to the geographic distance and then
identify the two fundamentally separated regimes in the
distribution.

\subsection{Number of Friends of a Particular User}
\label{SEC:Particular_User}

\begin{figure}[t!]
\centering{%
\subfigure[California]{%
\epsfxsize=0.49\textwidth \leavevmode \label{FIG:Friends_CA}
\epsffile{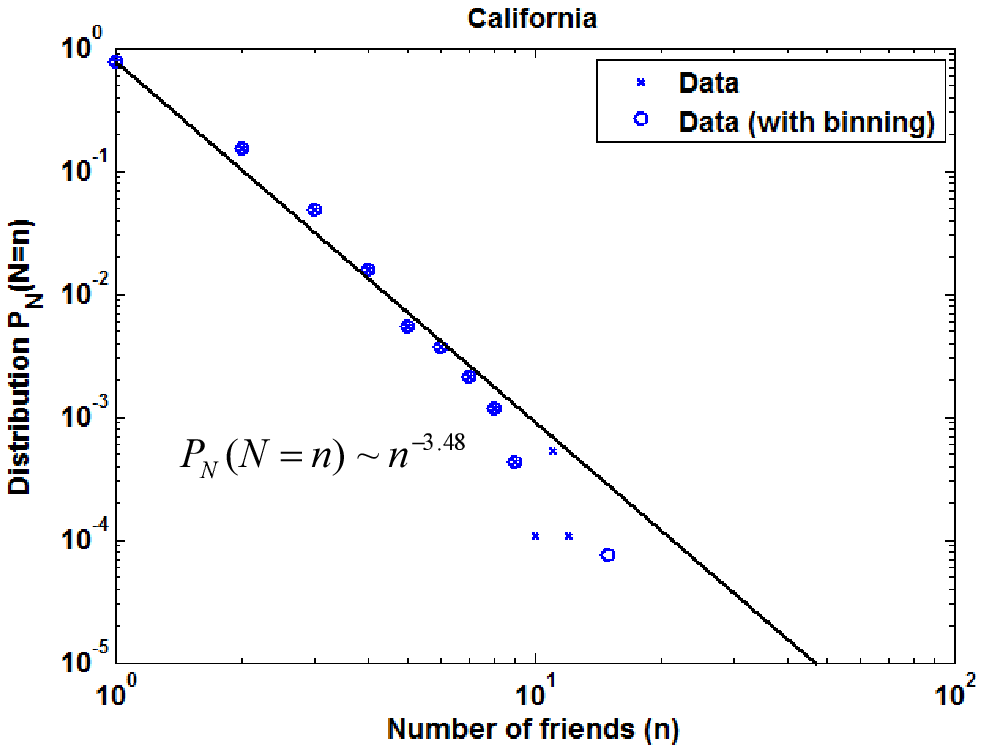}}\vspace{.0cm}
\subfigure[Los Angeles]{%
\epsfxsize=0.49\textwidth \leavevmode \label{FIG:Friends_LA}
\epsffile{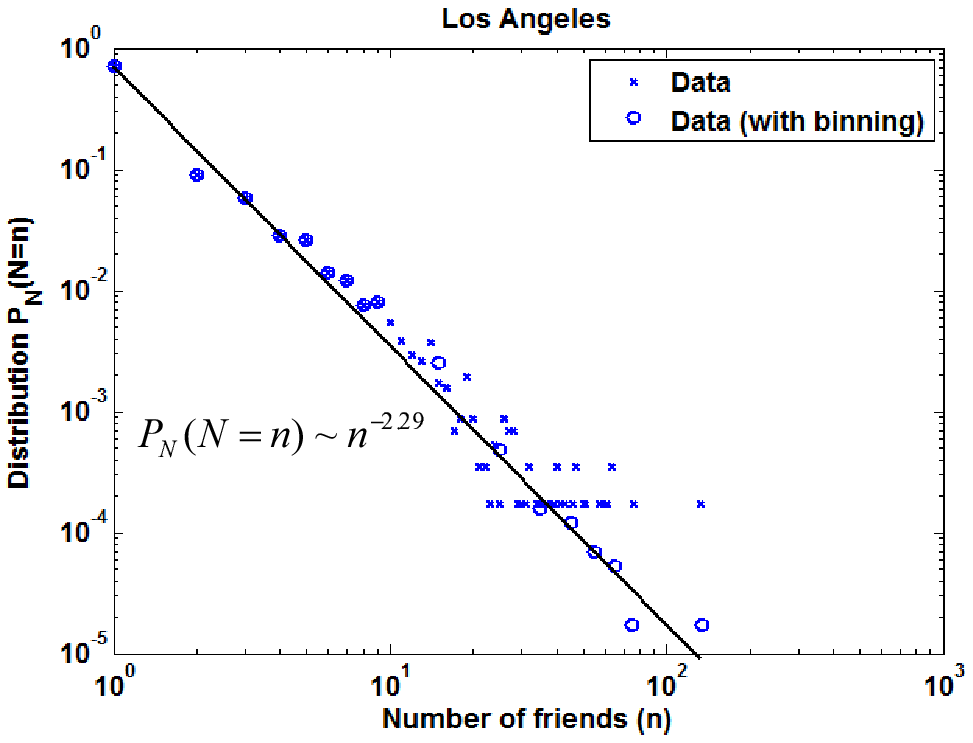}}\vspace{.0cm}
\subfigure[UK]{%
\epsfxsize=0.49\textwidth \leavevmode \label{FIG:Friends_UK}
\epsffile{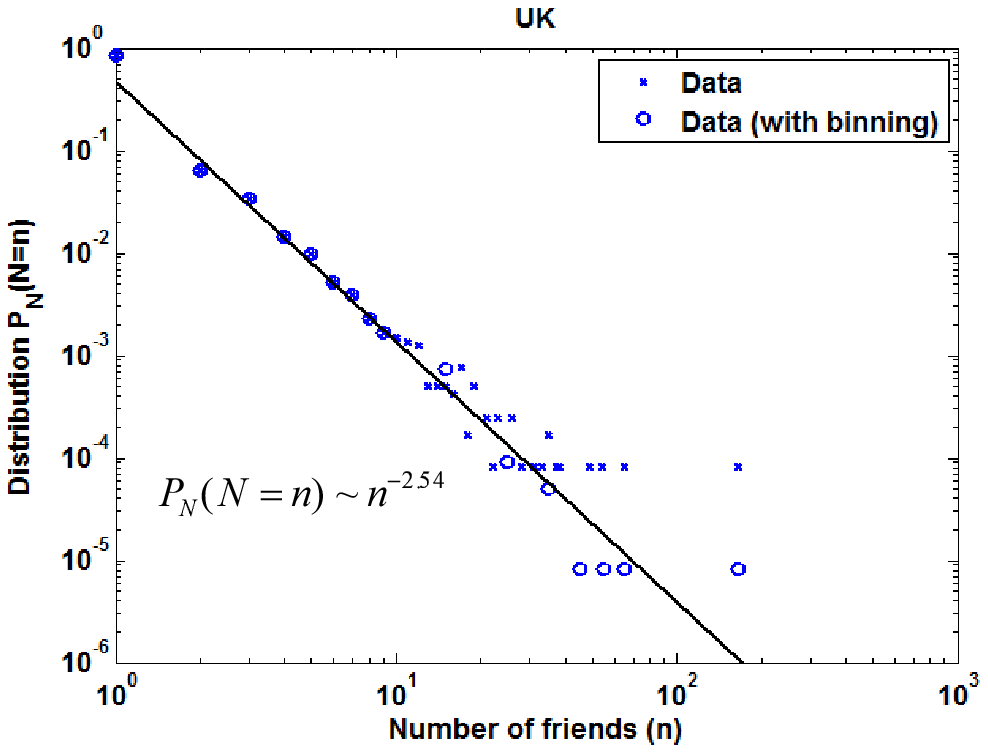}}\vspace{.0cm}
\subfigure[London]{%
\epsfxsize=0.49\textwidth \leavevmode \label{FIG:Friends_London}
\epsffile{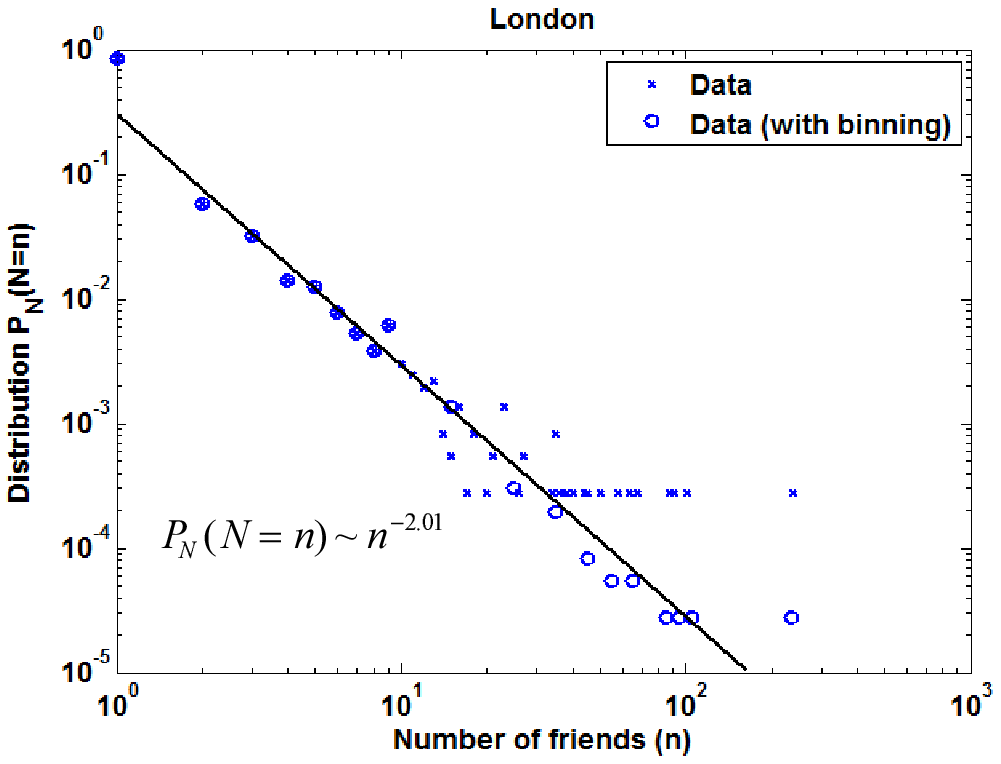}} } \caption{Probability distribution
$P_N(N=n)$ of the number of friends of a particular user (log-log
plot).} \label{FIG:Friends}
\end{figure}

We first find that the probability distribution $P_N(N=n)$ of the
number of friends for an individual, denoted by $n$, on Twitter
fits into a single power-law function $P_N(N=n)\sim n^{-\alpha}$
for $\alpha>0$. Figure~\ref{FIG:Friends} shows the log-log plot of
the distribution $P_N(N=n)$ obtained from empirical data,
logarithmically binned data, and fitting function, where the
fitting is applied to the binned data. As depicted in the figure,
statistical noise exists in the tail where the number of friends
is very large. Such noise can be eliminated by applying
logarithmic binning, which averages out the data that fall in
specific bins~\cite{Milojevic:10}.\footnote{It is also verified
that this binning procedure does not fundamentally change the
underlying power-law exponent of the distribution $P_N(N=n)$.} We
use the traditional least squares estimation to obtain the fitting
function. In TABLE~\ref{TABLE5}, the value of the exponent of
$P_N(N=n)$, $\alpha$, is summarized for each region. From
Figure~\ref{FIG:Friends} and TABLE~\ref{TABLE5}, the following
interesting comparisons are performed according to types of
regions:

\begin{table}[t!]
\renewcommand{\arraystretch}{1.1}
\begin{center}
\caption{The value of $\alpha$ for each region.}

 \begin{tabular}{|c||c|}
   \hline
    Region & $\alpha$   \\
 \hline
 \hline
    California & 3.48   \\
 \hline
  Los Angeles & 2.29  \\
 \hline
   UK & 2.54  \\
   \hline
   London & 2.01 \\
   \hline
\end{tabular}
\label{TABLE5}
\end{center}

\end{table}

\begin{itemize}
\item {\bf Comparison between the city-scale and
state-scale/country-scale results}: Figures~\ref{FIG:Friends_CA}
and~\ref{FIG:Friends_LA} illustrate that the exponent $\alpha$ is
3.48 and 2.29 in California and Los Angeles, respectively, which
implies that Twitter users in populous metropolitan areas are more
likely to contact a higher number of friends within a given period
(e.g., one month). From Figures~\ref{FIG:Friends_UK}
and~\ref{FIG:Friends_London}, the same trend is also observed by
comparing the results for the UK and London, with $\alpha$ values
of 2.54 and 2.01, respectively. That is, urban people are likely
to bilaterally interact with more friends by sending and receiving
directed geo-tagged mentions, compared on average to people in
larger regions that include local small towns.

\item {\bf Comparison between the results in the two cities (Los
Angeles and London)}: From Figures~\ref{FIG:Friends_LA}
and~\ref{FIG:Friends_London}, one can see that the exponent
$\alpha$ is 2.29 and 2.01 in Los Angeles and London, respectively.
This reveals that Twitter users in London tend to contact a
slightly higher number of friends within a given period, compared
to users in Los Angeles. There may be many explanations for this
phenomenon, including that i) London is one of the world's most
famous tourist destinations, which would attract relatively more
visitors to use Twitter to send/receive direct mentions to/from
their friends in the city and ii) London has a relatively higher
population density than that of Los Angeles (refer to TABLE~I for
more details).
\end{itemize}

\subsection{Number of Friends With Respect to Distance}

%
%
%
%
%

The most interesting characteristic in friendship degrees is how
friends of a user are distributed with respect to the geographic
distance between the Twitter user and his/her friend. In this
subsection, similarly as
in~\cite{Liben-Nowell:05,Kaltenbrunner:12}, we also verify whether
Twitter users establish more relationships with friends who are
living in geographic proximity to each other. As mentioned before,
in our experiments, we use geo-tagged mentions to identify the
location information of a user when he/she sent a mention to
his/her friend. To detect his/her friend's location, we then
observe {\em replied} geo-tagged mentions that were sent at the
next closest time. Using these bidirectional mentions, we
characterize the probability distribution $P_D(D=d)$ of the number
of friends according to the distance $d$, where $d$ [km] is the
geographic distance between a user and his/her friend.

Unlike the earlier work in~\cite{Liben-Nowell:05}, the
heterogeneous shape of $P_D(D=d)$ for the entire interval cannot
be captured by a single commonly-used statistical function such as
a homogeneous power-law using the approach of parametric fitting.
Interestingly, as our main result, we observe that for the
distance $d\in[d_{\text{min}},d_{\text{max}}]$, $P_D(D=d)$ can be
described as a {\em double power-law} distribution, which is given
as:

\begin{figure}[t!]
\centering{%
\subfigure[California]{%
\epsfxsize=0.49\textwidth \leavevmode \label{FIG:Distance_CA}
\epsffile{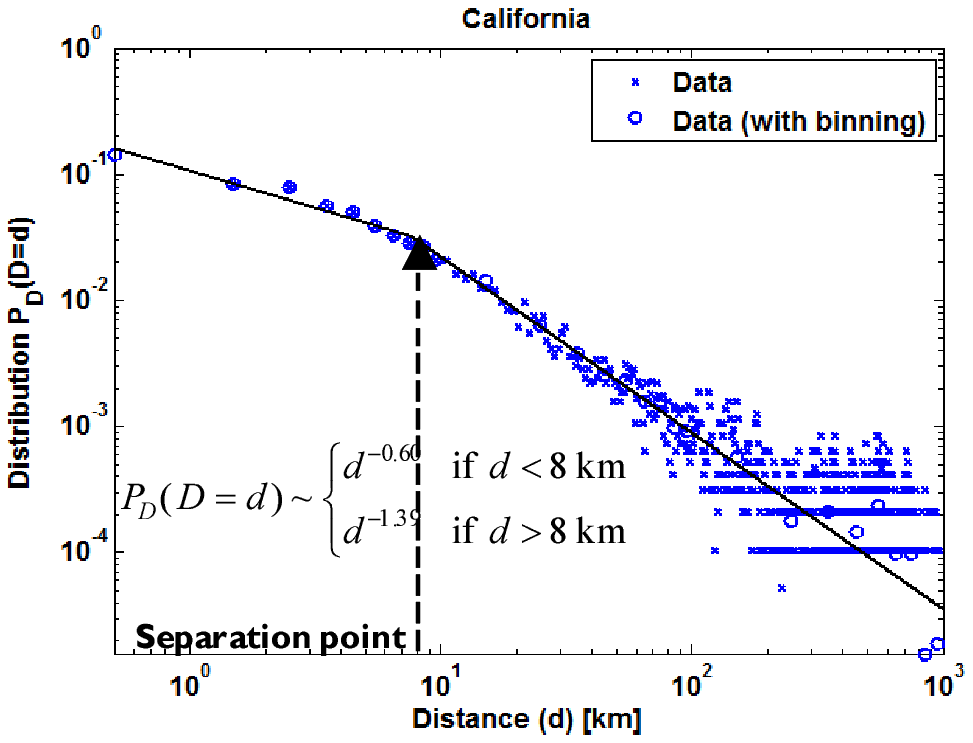}}\vspace{.0cm}
\subfigure[Los Angeles]{%
\epsfxsize=0.49\textwidth \leavevmode \label{FIG:Distance_LA}
\epsffile{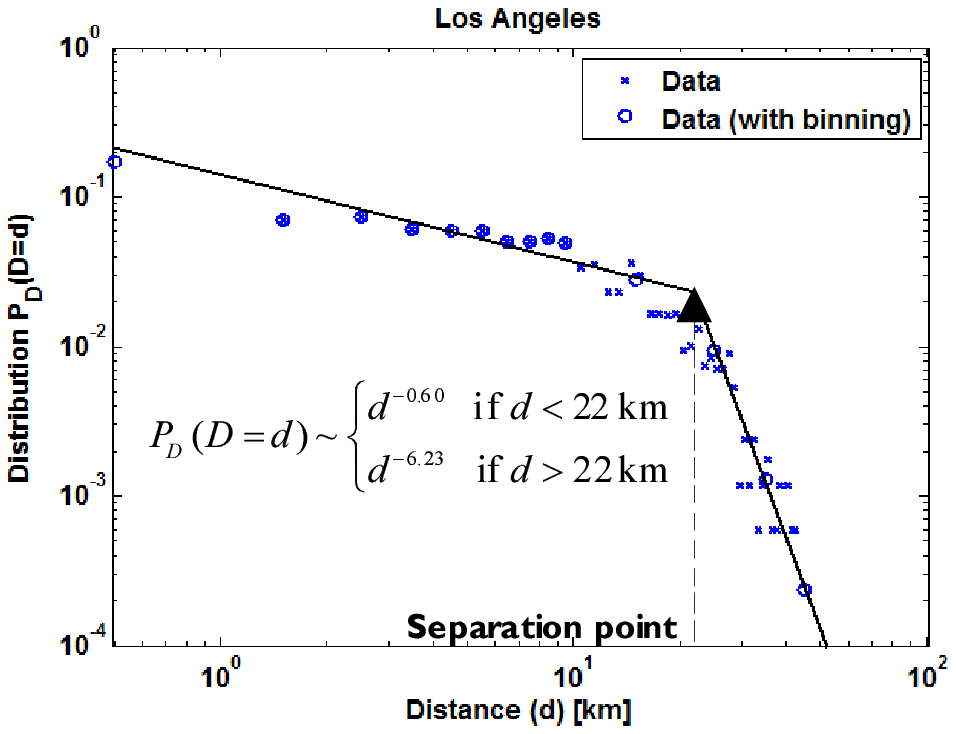}}\vspace{.0cm}
\subfigure[UK]{%
\epsfxsize=0.49\textwidth \leavevmode \label{FIG:Distance_UK}
\epsffile{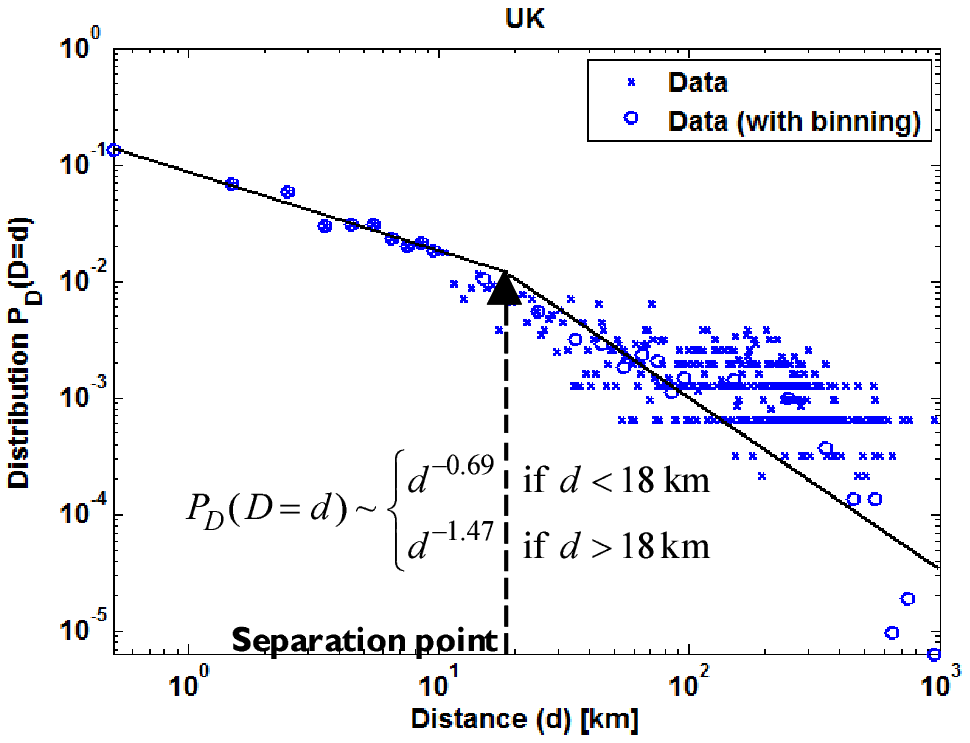}}\vspace{.0cm}
\subfigure[London]{%
\epsfxsize=0.49\textwidth \leavevmode \label{FIG:Distance_London}
\epsffile{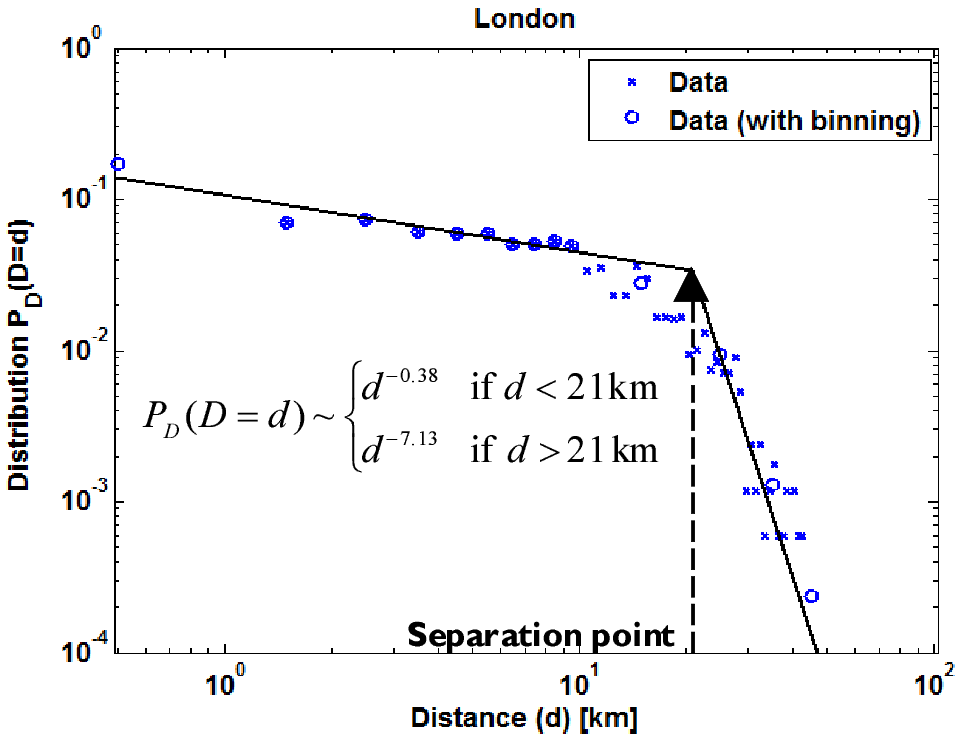}} } \caption{Probability
distribution $P_D(D=d)$ of the number of friends with respect to
distance (log-log plot).} \label{FIG:Distances}
\end{figure}

\begin{eqnarray}
P_D(D=d)\sim \left\{\begin{array}{lll} d^{-\gamma_1} &\textrm{if
$d_{\text{min}}\le d<d_s$ (intra-city regime)}
\\ d^{-\gamma_2} &\textrm{if $d_s\le d\le d_{\text{max}}$ (inter-city regime),}
\end{array}\right. \label{EQ:p_D}
\end{eqnarray}
where $\gamma_1$ and $\gamma_2$ denote the exponents for each
individual power-law and $d_s$ is the {\em separation point}. This
finding indicates that the friendship degree can be composed of
two {\em separate regimes} characterized by two different
power-laws, termed the {\em intra-city} and {\em inter-city}
regimes. Figure~\ref{FIG:Distances} shows the log-log plot of the
distribution $P_D(D=d)$ from empirical data, logarithmically
binned data, and fitting function, where the fitting is applied to
the binned data. As in Section~\ref{SEC:Particular_User}, we also
use the traditional least squares estimation to obtain the fitting
function.\footnote{Using maximum likelihood estimation to fit a
mixture function (e.g., a double power-law function) is not easy
to implement and the performance of mixture functions has not been
well understood.} In TABLE~\ref{TABLE6}, the value of the
exponents of $P_N(N=n)$, $\gamma_1$ and $\gamma_2$, is summarized
for each region.

Unlike the earlier studies
in~\cite{Liben-Nowell:05,Kaltenbrunner:12} that do not capture the
friendship patterns in the intra-city regime, our analysis
exhibits two distinguishable features with respect to distance.
More specifically, in each regime, the following interesting
observations are made:

\begin{table}[t!]
\renewcommand{\arraystretch}{1.1}
\begin{center}
\caption{The value of $\gamma_1$ and $\gamma_2$ for each region.}

 \begin{tabular}{|c||c|c|}
   \hline
    Region & $\gamma_1$ & $\gamma_2$   \\
 \hline
 \hline
    California & 0.60 & 1.39   \\
 \hline
  Los Angeles & 0.60 & 6.23  \\
 \hline
   UK & 0.69 & 1.47  \\
   \hline
   London & 0.38 & 7.13 \\
   \hline
\end{tabular}
\label{TABLE6}
\end{center}

\end{table}

\begin{itemize}
\item In the intra-city regime, the distribution $P_D(D=d)$ decays
slowly with distance $d$, which means that geographic proximity
weakly affects the number of intra-city friends with which one
user interacts. That is, in this regime, the geographic distance
is less relevant for determining the number of friends. This
finding reveals that more active Twitter users tend to
preferentially interact over {\em short-distance} connections.

\item In the inter-city regime, $P_D(D=d)$ depends strongly on the
geographic distance, where there exists a sharp transition in the
distribution $P_D(D=d)$ beyond the separation point $d_s$. Thus,
{\em long-distance} communication is made occasionally.
\end{itemize}

The above argument stems from the fact that the separation point
$d_s$ is closely related to the length and width of the city in
which a user resides. From these observations, we may conclude
that within a given period, the individual is much more likely to
contact online mostly friends who are in location-based
communities that range from the local neighborhood, suburb,
village, or town up to the city level. In addition, the following
interesting comparisons are performed according to types of
regions:

\begin{itemize}
\item {\bf Comparison between the city-scale and
state-scale/country-scale results}: We observe that the separation
point $d_s$ in populous metropolitan areas is much greater than
that in larger regions that include local small towns (such as at
the state or country level). For example, from
Figures~\ref{FIG:Distance_CA} and~\ref{FIG:Distance_LA}, we see
that $d_s$ is approximates 8 km and 22 km in California and Los
Angeles, respectively. From Figures~\ref{FIG:Distance_UK}
and~\ref{FIG:Distance_London}, the same trend is observed by
comparing the results for the UK and London (18 km and 21 km,
respectively). This finding reveals that Twitter users in populous
metropolitan areas (e.g., Los Angeles and London) have a stronger
tendency to contact friends on Twitter who are geographically away
from their location (i.e., interacting over long-distance
connections). This is because the average size (referred to as the
land area) of the considered metropolitan cities is relatively
bigger than that of cities in larger regions including small
towns. Furthermore, it is seen that the exponent in the inter-city
regimes (i.e., $\gamma_2$) in metropolitan areas is significantly
higher than that in larger regions. Unlike the
state-scale/country-scale results, this finding implies that the
distribution $P_D(D=d)$ sharply drops off beyond $d_s$ in huge
metropolitan areas.

\item {\bf Comparison between the results in the two cities (Los
Angeles and London)}: From Figures~\ref{FIG:Distance_LA}
and~\ref{FIG:Distance_London}, one can see that $\gamma_1$ is 0.60
and 0.38 and $\gamma_2$ is 6.23 and 7.13 in Los Angeles and
London, respectively. Thus, in the intra-city regime, the
geographic distance is less relevant in London for determining the
number of friends. However, in the inter-city regime, the
distribution $P_D(D=d)$ in London shows a bit steeper decline.
\end{itemize}

Our geo-tagged Twitter data provides position resolution at up to
10 meters, compared to the typical city-scale resolution in
previous studies on
friendship~\cite{Liben-Nowell:05,Kaltenbrunner:12}, thus allowing
much more fine-grained validation of these heterogeneous behaviors
in terms of distance.

\section{Concluding Remarks}

The present work has developed a novel framework for analyzing the
degree of bidirectional online friendship via Twitter, while not
only utilizing geo-tagged mentions but also introducing a
definition of bidirectional friendship. To provide analysis
results, we first introduced two new algorithms, the first for
counting friends and the second for a two-stage distance
estimation algorithm. We verified that the homogeneous power-law
model, also known as Zipf's law, holds on Twitter in terms of the
number of friends of one user. More interestingly, we
comprehensively demonstrated that the number of friends according
to geographic distance follows a double power-law distribution, or
equivalently, a double Pareto law distribution, where there exists
a strict separation point in distance that distinguishes the
intra-city regime from the inter-city regime. Our analysis sheds
light on a new understanding of social interaction/relationships
online with regard to small-scale space as well as large-scale
space.

Characterization of the degree of friendship in space along with a
greater variety of city/state/country-scale data on Twitter
remains for future work. Suggestions for further research in this
area also include analyzing a new friendship in the temporal
domain (time) by utilizing geo-located Twitter.



\begin{thebibliography}{10}
\providecommand{\url}[1]{#1} \csname url@rmstyle\endcsname
\providecommand{\newblock}{\relax}
\providecommand{\bibinfo}[2]{#2}
\providecommand\BIBentrySTDinterwordspacing{\spaceskip=0pt\relax}
\providecommand\BIBentryALTinterwordstretchfactor{4}
\providecommand\BIBentryALTinterwordspacing{\spaceskip=\fontdimen2\font
plus \BIBentryALTinterwordstretchfactor\fontdimen3\font minus
  \fontdimen4\font\relax}
\providecommand\BIBforeignlanguage[2]{{%
\expandafter\ifx\csname l@#1\endcsname\relax
\typeout{** WARNING: IEEEtran.bst: No hyphenation pattern has been}%
\typeout{** loaded for the language `#1'. Using the pattern for}%
\typeout{** the default language instead.}%
\else \language=\csname l@#1\endcsname \fi #2}}

\bibitem{Wilson:09}
Wilson C, Boe B, Sala A, Puttaswamy KPN, and Zhao BY. User
interactions in social networks and their implications. In: {\em
Proceedings of the 4th ACM European Conference on Computer Systems
(EuroSys'09)}, Nuremberg, Germany, March/April 2009, pp. 205--218.

\bibitem{Kwak:10}
Kwak H, Lee C, Park H, and Moon S. What is Twitter, a social
network or a news media?. In: {\em Proceedings of the 19th
International World Wide Web Conference (WWW2010)}, Raleigh, NC
USA, April 2010, pp. 591--600.

\bibitem{Viswanath:09}
Viswanath B, Mislove A, Cha M, and Gummadi KP. On the evolution of
user interaction in Facebook. In: {\em Proceedings of the 2nd ACM
Workshop on Online Social Networks (WOSN2009)}, Barcelona, Spain,
August 2009, pp. 37--42.

\bibitem{Mislove:08}
Mislove A, Koppula HS, Gummadi KP, Druschel P, and Bhattacharjee
B. Growth of the Flickr social network. In: {\em Proceedings of
the 1st ACM Workshop on Online Social Networks (WOSN2008)},
Seattle, WA USA, August 2008, pp. 25--30.

\bibitem{Chen:14}
Chen Y, Zhuang C, Cao Q, and Hui P. Understanding cross-site
linking in online social networks. In: {\em Proceedings of the 8th
ACM Workshop on Social Network Mining and Analysis (SNAKDD2014)},
New York City, NY USA, August 2014.

\bibitem{Watts:98}
Watts DJ and Strogatz SH. Collective dynamics of `small-world'
networks. {\em Nature} 1998; 393:440--442.

\bibitem{Svenson:06}
Svenson P. Complex networks and social network analysis in
information fusion. In: {\em Proceedings of the 9th International
Conference on Information Fusion (Fusion2006)}, Florence, Italy,
July 2006, pp. 1--7.

\bibitem{Liben-Nowell:05}
Liben-Nowell D, Novak J, Kumar R, Raghavan P, and Tomkins A.
Geographic routing in social networks. {\em Proceedings of the
National Academy of Sciences of the United States of America
(PNAS)} 2005; 102: 11623--11628.

\bibitem{Kaltenbrunner:12}
Kaltenbrunner A, Scellato S, Volkovich Y, Laniado D, Currie D,
Jutemar EJ, and Mascolo C. Far from the eyes, close on the web:
Impact of geographic distance on online social interactions. In:
{\em Proceedings of the 5th ACM Workshop on Online Social Networks
(WOSN'12)}, Helsinki, Finland, August 2012, pp. 19--24.

\bibitem{Jurdak:13}
Jurdak R, Corke P, Cotillon A, Dharman D, Crossman C, and Salagnac
G. Energy-efficient localization: GPS duty cycling with radio
ranging. {\em ACM Transactions on Sensor Networks (TOSN)} 2013; 9:
A:1--A:32.

\bibitem{Takhteyev:12}
Takhteyev Y, Gruzd A, and Wellman B. Geography of Twitter
networks. {\em Social Networks} 2012; 34: 73--81.

\bibitem{Kulshrestha:12}
Kulshrestha J, Kooti F, Nikravesh A, and Gummadi KP. Geographic
dissection of the Twitter network. In: {\em Proceedings of the 6th
International AAAI Conference on Weblogs and Social Media
(ICWSM-12)}, Dublin, Ireland, June 2012, pp. 202--209.

\bibitem{Frias-Martinez:12}
Frias-Martinez V, Soto V, Hohwald H, and Frias-Martinez E.
Characterizing urban landscapes using geolocated tweets. In: {\em
Proceedings of the 4th ASE/IEEE International Conference on Social
Computing (SocialCom2012) and the 4th ASE/IEEE International
Conference on Privacy, Security, Risk and Trust (PASSAT2012)},
Amsterdam, The Netherlands, September 2012, pp. 239--248.

\bibitem{Alowibdi:14}
Alowibdi JS, Ghani S, and Mokbel MF. VacationFinder: A tool for
collecting, analyzing, and visualizing geotagged Twitter data to
find top vacation spots. In: {\em Proceedings of the 6th ACM
SIGSPATIAL International Workshop on Location-Based Social
Networks (LBSN2014)}, Dallas, TX USA, November 2014.

\bibitem{Lee:10}
Lee R and Sumiya K. Measuring geographical regularities of crowd
behaviors for Twitter-based geo-social event detection. In: {\em
Proceedings of the 2nd ACM SIGSPATIAL International Workshop on
Location-Based Social Networks (LBSN2010)}, San Jose, CA USA,
November 2010, pp. 1--10.

\bibitem{Hawelka:14}
Hawelka B, Sitko I, Beinat E, Sobolevsky S, Kazakopoulos P, and
Ratti C. Geo-located Twitter as proxy for global mobility
patterns. {\em Cartography and Geographic Information Science
(CaGIS)} 2014; 41: 260--271.

\bibitem{Jurdak}
Jurdak R, Zhao K, Liu J, AbouJaoude M, Cameron M, and Newth D.
Understanding human mobility from Twitter. Preprint, [Online].
Available: http://arxiv.org/abs/1412.2154.

\bibitem{Liu}
Liu J, Zhao K, Khan S, Cameron M, and Jurdak R. Multi-scale
population and mobility estimation with geo-tagged tweets.
Preprint, [Online]. Available: http://arxiv.org/abs/1412.0327.

\bibitem{Falcone:14}
Falcone D, Mascolo C, Comito C, Talia D, and Crowcroft J. What is
this place? Inferring place categories through user patterns
identification in geo-tagged tweets. In: {\em Proceedings of the
6th International Conference on Mobile Computing, Applications and
Services (MobiCASE2014)}, Austin, TX USA, November 2014.

\bibitem{Morstatter:13}
Morstatter F, Pfeffer J, Liu H, and Carley KM. Is the sample good
enough? Comparing data from Twitters' Streaming API with Twitter's
Firehose. In: {\em Proceedings of the 7th International AAAI
Conference on Weblogs and Social Media (ICWSM-13)}, Boston, MA
USA, July 2013, pp. 400--408.

\bibitem{Gonzalez:08}
Gonzalez MC, Hidalgo CA, and Barabasi AL. Understanding individual
human mobility patterns. {\em Nature} 2008; 453: 779--782.

\bibitem{Song:10}
Song C, Koren T, Wang P, and Barabasi AL. Modelling the scaling
properties of human mobility. {\em Nature Physics} 2010; 6;
818--823.

\bibitem{Jiang:13}
Jiang S, Fiore GA, Yang Y, Ferreira, Jr. J, Frazzoli E, and
Gonzalez MC. A review of urban computing for mobile phone traces:
Current methods, challenges and opportunities. In: {\em
Proceedings of the 2nd ACM SIGKDD International Workshop on Urban
Computing (UrbComp2013)}, Chicago, IL USA, August 2013.

\bibitem{Wang:11}
Wang D, Pedreschi D, Song C, Giannotti F, and Barabasi A-L. Human
mobility, social ties, and link prediction. In: {\em Proceedings
of the 17th ACM SIGKDD International Conference on Knowledge
Discovery and Data Mining (KDD2011)}, San Diego, CA USA, August
2011, pp. 1100--1108.

\bibitem{Rhee:11}
Rhee I, Shin M, Hong S, Lee K, and Chong S. In the Levy-walk
nature of human mobility. {\em IEEE/ACM Transactions on Networking
(TON)} 2011; 19: 630--643.

\bibitem{Chaintreau:07}
Chaintreau A, Hui P, Crowcroft J, Diot C, Gass R, and Scott J.
Impact of human mobility on opportunistic forwarding algorithms.
{\em IEEE Transactions on Mobile Computing (TMC)} 2007; 6:
606--620.

\bibitem{Hui:08}
Hui P and Crowcroft J. Human mobility models and opportunistic
communication system design. {\em Philosophical Transactions of
The Royal Society A: Mathematical, Physical and Engineering
Sciences} 2008; 366: 2005--2016.

\bibitem{Cattuto:10}
Cattuto C, Van den Broeck W, Barrat A, Colizza V, Pinton J-F, and
Vespignani A. Dynamics of person-to-person interactions from
distributed RFID sensor networks. {\em PLOS ONE} 2010; 5: e11596.

\bibitem{Cho:11}
Cho E, Myers SA, and Leskovec J. Friendship and mobility: User
movement in location-based social networks. In: {\em Proceedings
of the 17th ACM SIGKDD International Conference on Knowledge
Discovery and Data Mining (KDD2011)}, San Diego, CA USA, August
2011, pp. 1082--1090.

\bibitem{Backstrom:10}
Backstrom L, Sun E, and Marlow C. Find me if you can: Improving
geographical prediction with social and spatial proximity. In:
{\em Proceedings of the 19th International World Wide Web
Conference (WWW2010)}, Raleigh, NC USA, April 2010, pp. 61--70.

\bibitem{Giannakidou:08}
Giannakidou E, Koutsonikola V, and Vakali A. Co-clustering tags
and social data sources. In: {\em Proceedings of the 9th
International Conference on Web-Age Information Management
(WAIM2008)}, Zhangjiajie, China, July 2008, pp. 317--324.

\bibitem{Nguyen:15}
Nguyen TT and Jung JJ. Exploiting geotagged resources to spatial
ranking by extending HITS algorithm. {\em Computer Science and
Information Systems (ComSIS)} 2015; 12: 185--201.

\bibitem{Manning:99}
Manning C and Schutze H. {\em Foundations of statistical natural
language processing}. Cambridge, MA: MIT Press, 1999.

\bibitem{Newman:05}
Newman MEJ. Power laws, Pareto distributions and Zipf's law. {\em
Contemporary Physics} 2005; 46: 323--351.

\bibitem{Reed:03}
Reed WJ. The Pareto law of income--an explanation and an
extension. {\em Physica A} 2003; 319: 469--486.

\bibitem{Morstatter:14}
Morstatter F, Pfeffer J, and Liu H, When is it biased? Assessing
the representativeness of Twitter's Streaming API. In: {\em
Proceedings of the 23rd International World Wide Web Conference
(WWW2013)}, Seoul, Korea, April 2014, pp. 555--556.

\bibitem{Hodas:13}
Hodas NO, Kooti F, and Lerman K. Friendship paradox redux: Your
friends are more interesting than you. In: {\em Proceedings of the
7th International AAAI Conference on Weblogs and Social Media
(ICWSM-13)}, Boston, MA USA, July 2013, pp. 1--8.

\bibitem{Bastos:12}
Bastos MT, Travitzki R, and Puschmann C. What sticks with whom?
Twitter follower-followee networks and news classification. In:
{\em Proceedings of the 6th International AAAI Conference on
Weblogs and Social Media (ICWSM-12) Workshop on the Potential of
Social Media Tools and Data for Journalists in the News Media
Industry}, Dublin, Ireland, June 2012, pp. 6--13.

\bibitem{Ugander}
Ugander J, Karrer B, Backstrom L, and Marlow C. The anatomy of the
Facebook social graph. Preprint, [Online]. Available:
http://arxiv.org/abs/1111.4503.

\bibitem{Huang:13}
Huang Y, Shen C, and Contractor NS. Distance matters: Exploring
proximity and homophily in virtual world networks. {\em Decision
Support Systems} 2013; 55: 969--977.

\bibitem{Ennis:13}
Ennis A, Chen L, Nugent C, Ioannidis G, and Stan A. High level
geospatial information discovery and fusion for geocoded
multimedia. {\em International Journal of Pervasive Computing and
Communications (IJPCC)} 2013; 9: 367--382.

\bibitem{Sinnott:84}
Sinnott RW. Virtues of the haversine. {\em Sky and Telescope}
1984; 68: 158.

\bibitem{Milojevic:10}
Milojevic S. Power-law distributions in information science:
Making the case for logarithmic binning. {\em Journal of the
American Society for Information Science and Technology (JASIST)}
2010; 61: 2417--2425.

\end{thebibliography}
\end{document}